\documentclass[twocolumn,prd,showpacs,superscriptaddress,groupedaddress,preprintnumbers]{revtex4-1}
\usepackage{graphicx,amsfonts,amsmath,amssymb,epsfig,color, mathrsfs,latexsym,url,wasysym,float}
\newcommand{\be}{\begin{equation}}
\newcommand{\ee}{\end{equation}}
\newcommand{\nn}{\nonumber}
\newcommand{\ba}{\begin{eqnarray}}
\newcommand{\ea}{\end{eqnarray}}
\newcommand{\mpl}{m_{\rm Pl}}
\begin{document}

\title{When Higgs Meets Starobinsky in the Early Universe}
 
\author{Mahdi Torabian$^*$}
\affiliation{School of Particles and Accelerators, Institute for Research in Fundamental Sciences (IPM), 19395-5531, Tehran, Iran}

\begin{abstract}
The measurement of the Higgs mass at the LHC has confirmed that the Standard Model electroweak vacuum is a shallow local minimum and is not absolutely stable. In addition to a probable unacceptably fast tunneling to the deep true minimum, it is not clear how the observable present-day vacuum could be reached from the early Universe particularly following inflation. In this note it is shown that these problems can be alleviated if the Higgs field is non-minimally coupled to a higher-curvature theory of gravity which is effective in deriving inflation a la Starobinsky. Moreover, it implies that the Higgs self-coupling could be enhanced and have an observable effect at the next generation of particle colliders.  
\end{abstract}
\preprint{IPM/PA-381}
\maketitle

\subsection*{Introduction} 
The great achievement of the LHC was the discovery of the Higgs particle of mass 
$m_{h} = 125.15\pm0.24$  GeV \cite{Aad:2012tfa,Chatrchyan:2012ufa}. It completed the scalar sector of the Standard Model of particle physics with all parameters determined. In particular, the Higgs quartic self-coupling parameter is deduced at the electroweak scale to be $\lambda(m_{\rm EW})  \approx 0.13$. 
This parameter is the only parameter of the Standard Model which is not multiplicatively renormalized. 
Given the central value of top quark mass $m_t = 173.2\pm0.9$ GeV \cite{ATLAS:2014wva}, its beta-function $\beta_\lambda$ at low energy is chiefly dominated by the top Yukawa coupling and so is negative. 
The Standard Model is a renormalizable theory and thus one could in principle extrapolate it to an arbitrary high energy and make predictions. If no new physics intervenes, the effective Higgs potential can be computed at desired loop orders. The quartic coupling is monotonically decreasing, vanishes at an intermediate energy of about $10^{10}$ GeV and subsequently turns negative \cite{Degrassi:2012ry,Buttazzo:2013uya}
. At high energy the gauge interactions take over and make the beta function positive. Then, the quartic coupling is increasing as it develops a new minimum which will be the global one. The potential is plotted in Figure 1.
\begin{center}\begin{figure}[h]\includegraphics[scale=.34]{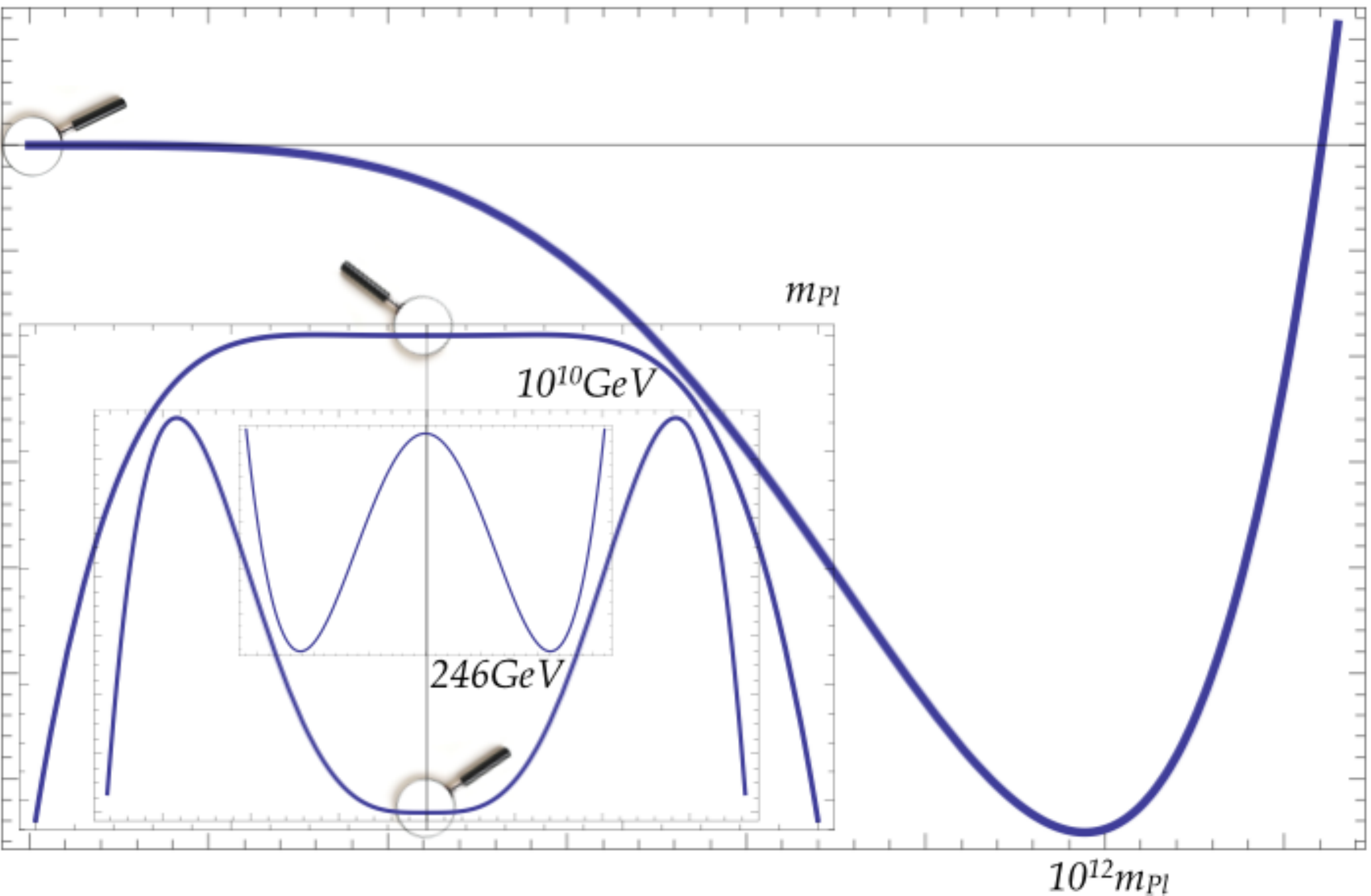}\caption{The 2-loop effective potential for the Higgs $V_{\rm eff}(\varphi) = \frac{1}{2}m^2(\mu)\varphi^2 + \frac{1}{4}\lambda(\mu)\varphi^4$ extrapolated to an arbitrary high scale. Magnified parts of the potential are shown as an ordered series of insets. The stability is disfavored at 98\% C.L.}\end{figure}\end{center}

The global minimum, located at tens of $\mpl$, is deeply inside the quantum gravity regime. This picture could radically be altered if one sensibly includes quantum gravity effects at high energies. Not understanding that thoroughly, it is reasonable to abandon the naive extrapolation to an arbitrary high energy and limit to the Planck scale. Thus, the potential would basically be seen ill as unbounded from below. The electroweak vacuum is a local one and barrier separating it from the deep well is extremely small. If one naively computes the tunneling rate between the vacua, one finds that life-time of the present-day electroweak vacuum is greater than the age of the Universe and thus the vacuum is {\it metastable} \cite{Sher:1988mj}. However, if one includes higher dimensional Planck-suppressed operators, the conclusion will be drastically changed and the electroweak vacuum turns {\it unstable} and quickly decays to true vacuum \cite{Branchina:2013jra,Branchina:2014usa,Branchina:2014rva}
. Consequently, due to a huge negative cosmological constant, it leads to a catastrophic gravitational collapse.

Furthermore, this picture of the Higgs potential raises problems in relation to the early Universe.
In order to end up in the energetically-disfavored present-day electroweak vacuum and prevent the Higgs from rolling down to the true minimum, a fine-tuning at level of {\it one} part in {\it a hundred million} in the Higgs value is needed \cite{Lebedev:2012sy,Bars:2013vba}. Moreover, it being granted that the Higgs is initially near the desired local minimum, it will not stick to that in the presence of Hubble-size quantum fluctuations during an intermediate to high scale inflation \cite{Ade:2013zuv,Ade:2014xna}.

New physics beyond the Standard Model could possibly stabilize the Higgs potential. However, excellent agreement of the Standard Model predictions with the experimental results puts tight constraints on new physics as it must have marginal effect on the electroweak fit. Moreover, generically new physics would inevitably introduce a naturalness problem to the scalar sector. 

Here it is argues that the stability problem can be alleviated without introducing unnaturalness if the Higgs field is non-minimally coupled to higher-curvature theory of gravity which account for the dynamics in the early Universe (see \cite{Kamada:2014ufa,Hook:2014uia,Herranen:2014cua,Hertzberg:2011rc,Hertzberg:2012zc,Onofrio:2014txa,Onofrio:2010zz,Fairbairn:2014zia} for related works). Basically, the idea is not to introduce new particles or interactions but to properly involve gravity at high energy scale.

\subsection*{The Higgs Sector Action}
The dynamics of the Higgs field which is non-minimally coupled to a higher-curvature theory of gravity is given by the following action
\be S=\int {\rm d}^4x(-g)^{1/2}\Big(\frac{1}{2} \mpl^2f(\varphi,R) - \frac{1}{2}g^{\mu\nu}\partial_\mu\varphi\partial_\nu\varphi -V(\varphi)\Big) ,\ee
where $\varphi^2 = 2 H^\dagger H$ and
\ba f(\varphi,R) &=&(1-\xi\mpl^{-2}\varphi^2) R + \alpha \mpl^{-2}R^2,\\V(\varphi) &=& \frac{1}{2}m^2\varphi^2 + \frac{1}{4}\lambda\varphi^4.\ea
The non-minimal parameter $\xi$ could be zero at tree-level. However, being related to a dimension 4 operator, it is unstable to quantum fluctuations and would be generated at loop-level in a curved spacetime \cite{Callan:1970ze}. Moreover, the term quadratic in the Ricci scalar is the simplest generalization to General Relativity regarding higher-oder gravity. Although being a higher-derivative theory of gravity, it is free from Ostrogradski classical instability or the presence of spin-2 ghost (and also spin-0 ghost for positive $\alpha$) in the spectrum \cite{Stelle:1976gc}.  
Furthermore, this terms is needed for the renormalization of the effective potential. Namely, the presence of higher derivative terms is needed if one demands renormalizability. 

Through a Weyl transformation of the metric
\be g_{\mu\nu}^E = (1-\xi\mpl^{-2}\varphi^2 + 2\alpha \mpl^{-2}R) g_{\mu\nu} \equiv e^{\tilde\chi} g_{\mu\nu},\ee
one can move to the Einstein frame with the following action
\ba\label{Einstein-frame-action} S_E = \int {\rm d}^4x(-g_E)^{1/2}\Big(&&\frac{1}{2} \mpl^2R_E - \frac{1}{2}g_E^{\mu\nu}\partial_\mu\chi\partial_\nu\chi \cr - &&\frac{1}{2}e^{-\tilde\chi}g_E^{\mu\nu}\partial_\mu\varphi\partial_\nu\varphi  -V_E(\varphi,\chi)\Big)\!.\ \ \ea 
The term quadratic in the Ricci scalar introduces a new propagating scalar, {\it a.k.a.} Weyl scalar, in the Einstein frame. The canonically normalized dimensionful Weyl scalar field is $\chi = (3/2)^{1/2}\mpl\tilde\chi$. Note that the Weyl scalar has a canonical kinetic term while the Higgs scalar has a non-canonical one. In fact, the Weyl and the Higgs fields interact both via the kinetic term and the scalar potential. Finally, the mass parameter in the Jordan and the Planck mass in Einstein frame, which is shown symbolically the same, might have slight different value depending on the value of $\xi$. 

\subsection*{Stability Condition}
The scalar potential in the Einstein frame reads as
\ba\label{potential} V_E(\varphi,\chi)\!=\! e^{-2\tilde\chi}V(\varphi)\!+\! \frac{1}{8}\alpha^{-1}\mpl^4\big(1\!-\!e^{-\tilde\chi}(1\!-\!\xi\mpl^{-2}\varphi^2)\big)^2\!\nn \\ = \frac{1}{8\alpha}\big(1-e^{-\tilde\chi}\big)^2\mpl^4 \qquad\qquad\qquad\qquad\qquad\quad\ \\ + \frac{1}{4}e^{-2\tilde\chi}\Big[2m^2+(e^{\tilde\chi}-1)\mpl^2\xi/\alpha+ \big(\lambda+\xi^2/2\alpha\big)\varphi^2\Big]\varphi^2,\nonumber\ea

The second line in equation \eqref{potential} gives the pure Weyl potential which is flat for large $\chi$ values. When the potential is mainly dominated by that, it derives a Starobinsky-like $R^2$-inflation in the early Universe \cite{Starobinsky:1980te}. The Planck results on the CMB anisotropy ($A_s=2.23\pm0.16\times10^{-9}$ 68\% C.L.)  and the primordial gravitational waves ($r<0.11$ 95\% C.L.) \cite{Ade:2013zuv} constraint the free parameter $\alpha$ as
\be \alpha = (12\pi^2 r A_s)^{-1}\gtrsim 3.4\times 10^7. \ee
The simplest manifestation of the Starobinsky inflation predicts $r\approx2.5\times 10^{-3}$ and therefor $\alpha\approx 1.5\times 10^9$. However, modifications to the model predict larger $r$ and so smaller $\alpha$ works as well (see \cite{Ben-Dayan:2014isa} and \cite{Neupane:2014vwa}
). For completeness, although not quite relevant to the present study, there is also an upper bound on the value of $\alpha$ from gravitational experiments measuring Yukawa correction to the Newtonian potential as $\alpha\lesssim 10^{61}$ \cite{Hoyle:2004cw,Calmet:2008tn}.

From the last line in equation \eqref{potential} one finds that depending on the sign and the absolute value of $\xi$, different stable solutions can be obtained. 
If the value of  $\xi$ parameter is large enough such that at the Planck scale the following constraint is sarisfied
\be\label{condition-1} \lambda(\mpl)+ \xi(\mpl)^2/2\alpha \geq 0,\ee
then the Higgs quartic coupling is positive-definite. Given the value of the Higgs quartic coupling at the Planck scale $\lambda(\mpl)\approx -0.01$ the above constraint implies that (for $\alpha\approx10^9$)
\be \xi(\mpl)\gtrsim 4500\quad {\rm or}\quad \xi(\mpl)\lesssim -4500.\ee
If $\xi>0$, then the Higgs potential is convex for any value of the scalar fields. If $\xi<0$, then the stability can be reached only if the Higgs quadratic term is suppressed through the following constraint on the field values
\be \tilde\chi_0 \lesssim 2 \ln\tilde\varphi_0 + \ln(-\xi/2).\ee

Furthermore, an absolutely stable potential exists for smaller values of $\xi$ when condition \eqref{condition-1} is relaxed. In fact, the Weyl field needs to take large value so that the Higgs quadratic term takes over the quartic term and makes the potential convex. Thus, stable solution exists provided that $\xi>0$ and the following condition is met
\be \tilde\chi_0 \gtrsim 16.1 + 2 \ln\tilde\varphi_0 - \ln \xi,\ee
so that the Higgs field receives a huge effective mass.  

The parameter $\xi$ basically normalizes (defines) the 4-dimensional Planck mass in the Einstein frame. An upper limit exists only when it is negative as $\xi \lesssim 10^{32}$. Later, using collider bounds, it will be argued that a tighter constraint on its absolute value can be assumed.

\subsection*{Fields Evolution}
The evolution of the Higgs and the Weyl field in the early Universe in a homogenous/isotropic background 
\ba {\rm d}s^2 = -{\rm d}t^2 + a(t){\rm d}{\bf x},\quad
\varphi = \varphi(t),\quad \chi = \chi(t),\ea
is governed by the following equations of motion
\ba \ddot\chi + 3H\dot\chi + 6^{-1/2}e^{-\tilde\chi}\dot\varphi^2+V^E_\chi &=& 0,\\
\ddot\varphi + 3H\dot\varphi - (2/3)^{1/2}\mpl^{-1} \dot\chi\dot\varphi +V^E_\varphi &=& 0,\ea
subject to the following metric equations
\ba 3 H^2 \mpl^2 &=& \frac{1}{2}\dot\chi^2 + \frac{1}{2}e^{-\tilde\chi} \dot\varphi^2 + V_E,\\
-2 \dot H \mpl^2 &=& \dot\chi^2 + e^{-\tilde\chi} \dot\varphi^2.\ea
\begin{center}\begin{figure}[t!]\includegraphics[scale=.51]{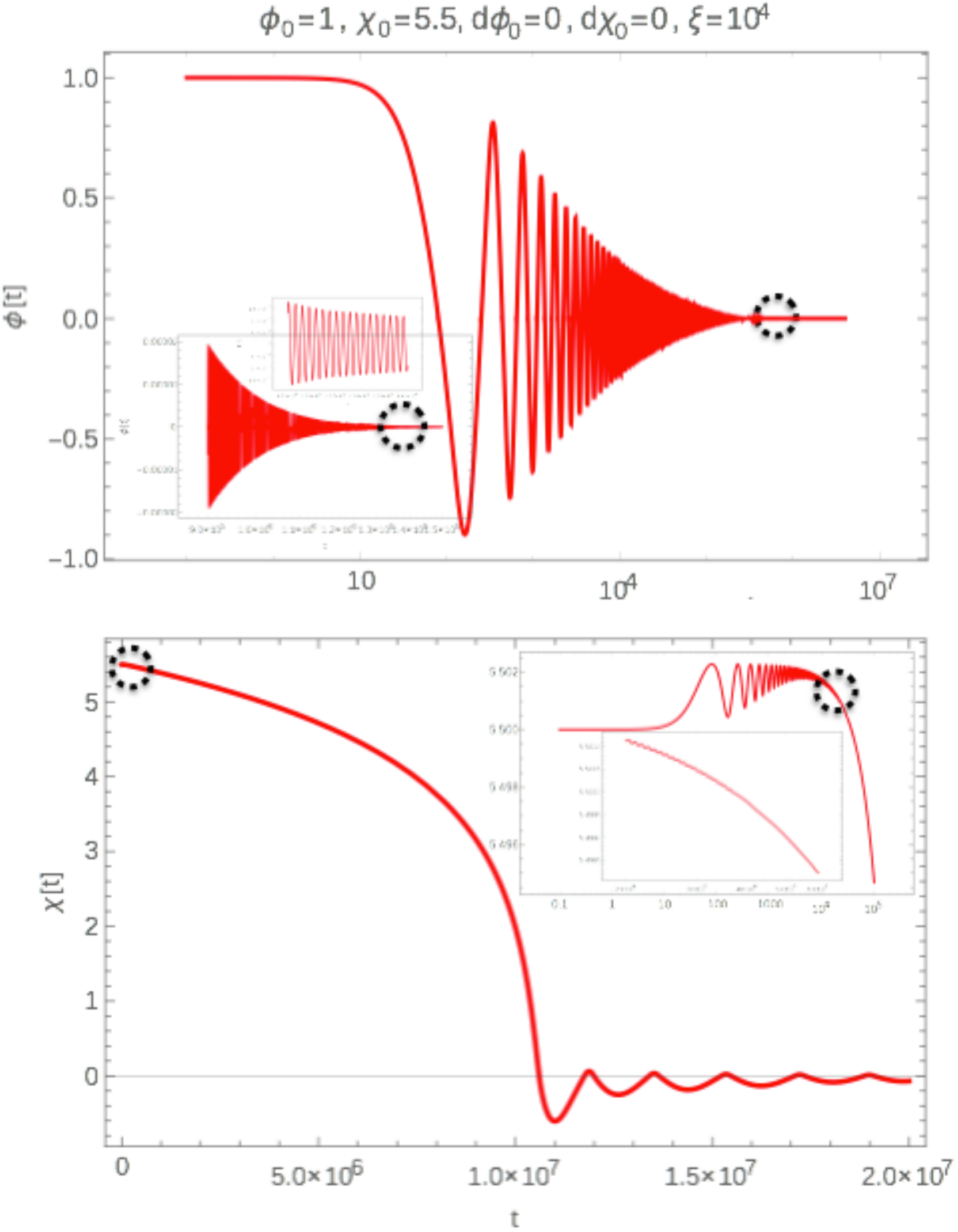}\caption{Time evolution of the Higgs (up) and the Weyl (down) fields. The initial conditions are mentioned on the top.}\end{figure}\end{center}
\begin{center}\begin{figure}[t!]\includegraphics[scale=.49]{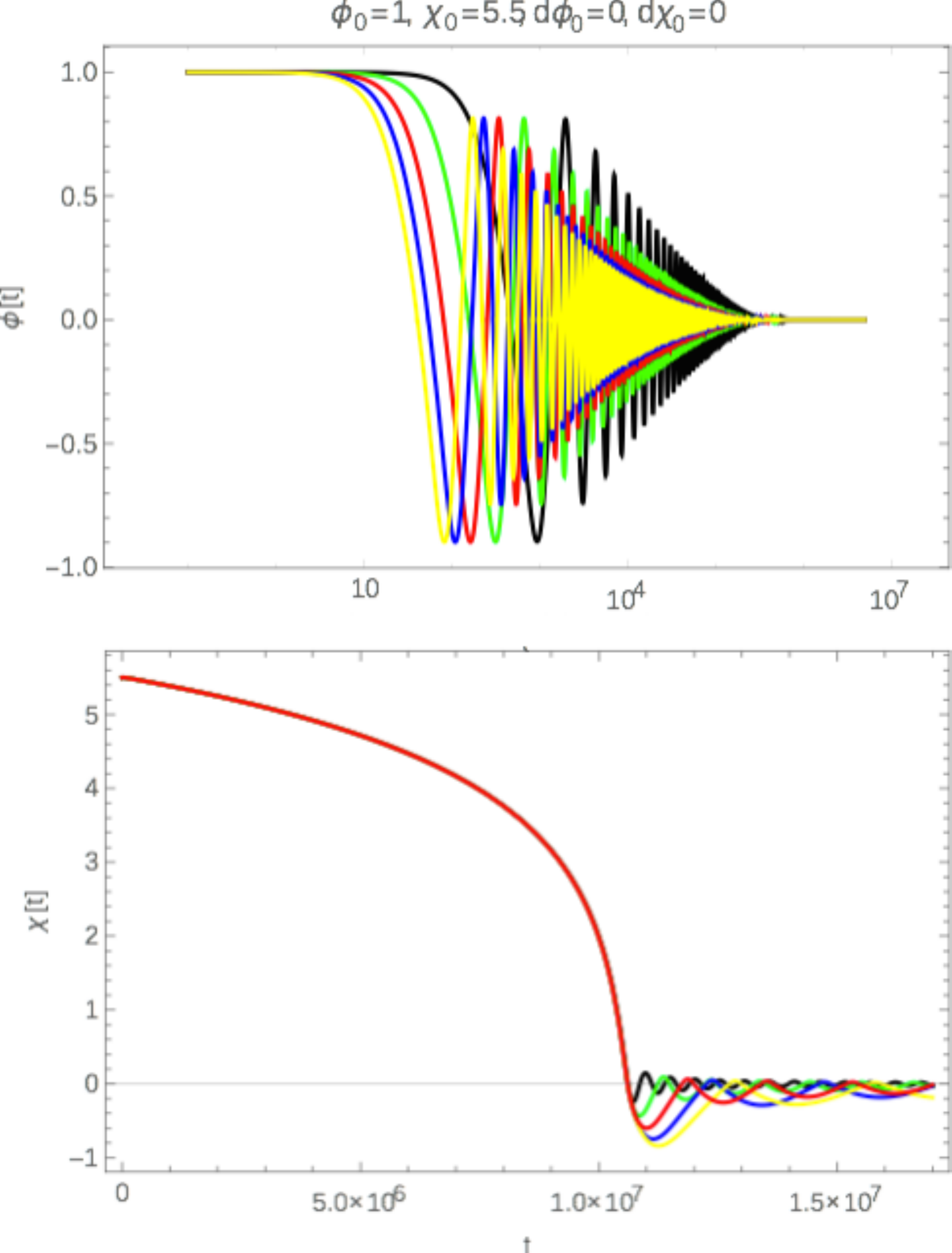}\caption{solutions for different $\xi$ parameter: $\xi$=1000(black), 5000(green), 10000(red), 15000(blue), 20000(yellow)}\end{figure}\end{center}
\vspace*{-14mm}
These are coupled second-order differential equations that can be solved by numerical methods. The solutions as scalar fields time evolution, in Planck units for field values and time, are plotted in Figure 2. 
The Higgs field initial value is of order one in Planck mass. Its initial velocity could also be chosen order one, however, it is found that it has insignificant qualitative effect on the solutions. On the other hand, the Weyl field initial conditions are chosen such that the Universe experiences at least 60 e-folds of exponential expansion.

The solutions are interpreted as follows. The Higgs and the Weyl fields are initially frozen for tens of Planck time until they commence harmonic oscillations about their local minima. Then, the Higgs field oscillates with large amplitude which is rapidly decreasing in time. Meanwhile, after a tiny jump in the value, the Weyl field oscillates with rather small amplitude which is slowly decreasing in time.  After about a hundred thousand Planck time the Weyl field slowly rolls down the potential from its initial value to smaller values while  the Higgs, whose value is reduced to $10^{-8}\mpl$, is still rapidly oscillating. 
This epoch lasts for about 60 Hubble times after which both fields oscillate about their minima and the Universe is filled by the Bose condensates of Higgs and Weyl particles. 
After many damped oscillations fields settle down in their minima near the origin. Finally, they decay and reheat the Universe through the Higgs decay products.  

The Hubble parameter is fairly constant in two time periods which can be interpreted as two inflationary epochs. The former (when $H\approx8.5\times10^{-3}\mpl$) could last for only few e-folds and the observable latter one (when $H\approx6.5\times10^{-6}\mpl$) lasts for about 60 e-folds. It is important to emphasis that by this time the Higgs field amplitude is less than $10^{-8}\mpl$ so it later evolves to the electroweak vacuum. 

Fields have slightly different evolution, although qualitatively the same, depending on the value of non-minimal coupling parameter. For completeness, this change behavior is plotted in Figure 3.

\subsection*{Electroweak Vacuum and the Higgs Coupling}
At late times, the Higgs field can be trapped to the local minima of its potential and spontaneously break the electroweak symmetry. The Weyl fields, as a result of the Higgs non-minimal coupling, falls into a non-trivial minima. The non-zero vacuum expectation values are 
\ba \varphi_0^2 &=& -m^2/\big(\lambda+\xi (m/\mpl)^2\big),\\
e^{\tilde\chi_0} &=& 1-\varphi_0^2\big(\xi-2\alpha(m/\mpl)^2\big)\mpl^{-2}.\ea
The later implies that the Higgs needs a field redefinition to make its kinetic term canonical in the action \eqref{Einstein-frame-action}
\be \varphi_c = e^{-\tilde\chi_0/2}\varphi.\ee
Phenomenologically, one is interested in the solution 
\ba \label{vev}\varphi_0 &\equiv& v \approx -m/\lambda^{1/2} \approx 246\ {\rm GeV}, \\ |\chi_0| &\approx& |\xi|\mpl^{-1} v^2 \approx |\xi| \times10^{-5}{\rm eV}.\ea

As a result of the Higgs-Weyl interaction in the potential the mass matrix is not diagonal. 
It can be easily diagonalized as
\be {\rm diag}(M^2, m_h^2) = R(-\theta)\begin{pmatrix}V_{\chi\chi}&V_{\chi\varphi_c}\\V_{\chi\varphi_c}&V_{\varphi_c\varphi_c}\end{pmatrix}R(\theta),\ee
where $R$ is the rotation matrix and the mixing angle is
\be \tan2\theta = 2V_{\chi\varphi_c}/(V_{\chi\chi}-V_{\varphi_c\varphi_c}).\ee
Moreover, the mass eigenstates are parametrized by $h$ and $H$ where
the {\it physical} Higgs field is
\be h = \cos\theta e^{-\tilde\chi_0/2}\varphi + \sin\theta\, \chi.\ee
Consequently, the coupling of the Higgs boson to the rest of SM particles is suppressed and it effectively decouples for large mixing angles. Thus, its production and decay rate will be affected. This modification has an observable effect at colliders in terms of  the global signal strength
\be \mu\equiv\sigma\cdot{\rm Br}/\sigma_{\rm SM}\cdot{\rm Br}_{\rm SM} = e^{-\tilde\chi}\cos^2\theta.\ee
The combined analysis of the ATLAS and CMS measurements implies $\mu= 1.07\pm0.18$  \cite{Aad:2012tfa,Chatrchyan:2012ufa,Atkins:2012yn}. 
Therefore, it implies that the mixing angle should be small and approximately $\theta\approx\sqrt{6}\mpl^{-1}\xi v\ll1$. Then
\be \mu \approx (1-\tilde\chi)\cdot(1-\theta^2)\approx 1-6\mpl^{-2}\xi^2v^2,\ee
and so this measurement excludes $|\xi|\gtrsim 10^{15}$ at $95\%$ C.L.


A tighter constraint can be found by analyzing the mass eigenvalues. The mass of the heaviest state $H$ is 
\be M \approx \mpl(6\alpha)^{-1/2} \approx 10^{13} {\rm GeV}.\ee
This is the mass of inflaton which is effectively decoupled from low-scale physics. 
The lightest mass eigenvalue, the mass of the physical Higgs particle, is found as 
\be m_h^2 = \bigg[2\lambda - \frac{\xi^2}{2\alpha}\frac{(2\mpl^{-1}\xi v)^2}{1+(2\mpl^{-1}\xi v)^2}\bigg]v^2 \equiv 2\lambda_{\rm eff} v^2.\ee 
The LHC historic measurement implies that
\be\label{mass} \lambda_{\rm eff}  \approxeq \lambda+ \mpl^{-2}\alpha^{-1}\xi^4v^2 \approx 0.13.\ee
Accordingly in contrast to the SM, the two measurements of $v$ and $m_h$ only fix two combinations of $m$, $\lambda$ and $\xi$ through \eqref{vev} and \eqref{mass}.
In the outer panel of Figure 4, $\lambda_{\rm eff}$ is plotted versus $\xi$ for different values of $\lambda$ from 0.13 to 0.2.  Experiment selects regions in the parameter space where each curve crosses the horizontal dotted line. 

On the other hand, the quartic self-coupling of the physical Higgs can be evaluated as follows
\be \lambda_{h} = \lambda\cos^4\theta \approx \lambda (1-2\theta^2) \approx \lambda \big(1-3(2\mpl^{-1}\xi v)^2\big).\ee
Then, using \eqref{mass} one finds that
\be \lambda_h \approx (0.13 + \mpl^{-2}\alpha^{-1}\xi^4v^2)(1-12\mpl^{-2}\xi^2v^2),\ee

Therefore, a prominent {\it prediction} of this mode is that the physical Higgs self-coupling is greater than what is inferred in the SM. The next generation of particle colliders are commissioned to directly measure the Higgs self-coupling in multi-Higgs processes. If a significant deviation from the SM values is observed, then new physics will be immediately need to account for the rest. The physical quartic coupling is plotted in the inner panel of Figure 4 versus $\xi$. It is interesting to note that perturbativity ({\it i.e.} $\lambda_h\lesssim 1$) imposes a tighter upper bound on the absolute value of the Higgs non-minimal coupling parameter as $|\xi| \lesssim 10^{10}$.

\begin{center}\begin{figure}[t!]\includegraphics[scale=.3]{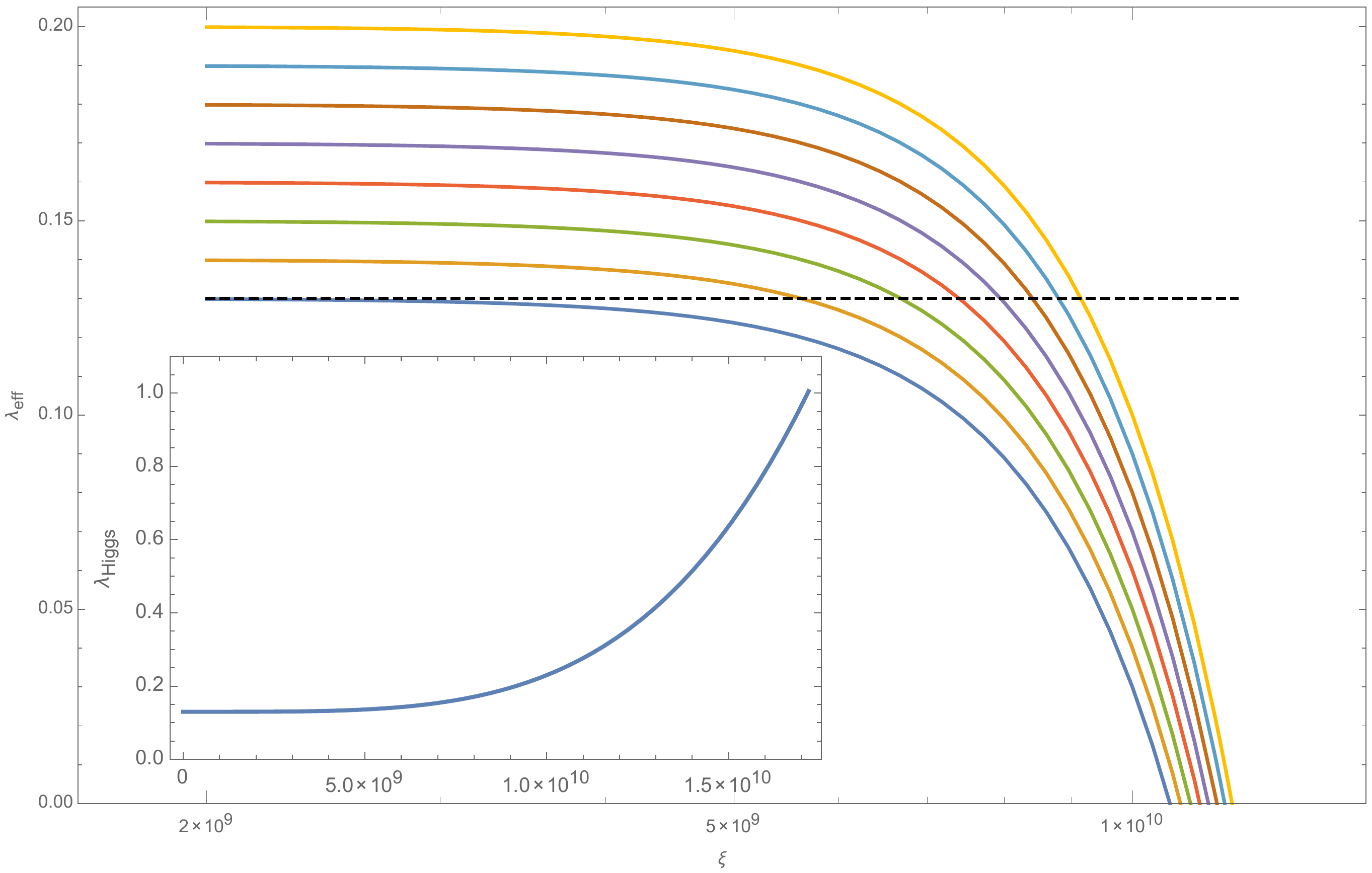}\vspace*{-5mm}\caption{The outer and inner panel respectively: $\lambda_{\rm eff}$ (for different $\lambda$) and $\lambda_h$ versus $\xi$.}\end{figure}\end{center}

\vspace*{-8mm}\subsection*{Conclusion}
In this note it has been argued that non-minimal coupling of the Higgs field to a higher-curvature theory of gravity  could alleviate problems associate with metastability or instability the electroweak vacuum. Essentially, pure gravitational effects, operative in deriving inflation, could be applied to account for the present-day electroweak minimum without a need for fine-tinning in the initial Higgs value in the early times. Besides, it implies that the Higgs self-coupling could be enhanced which would be observable at the next generation of colliders. 

\paragraph*{Acknowledgments}
I would like to thank Collaborative Research Center SFS676 at the University of Hamburg and ICTP, Trieste for hospitality during the initial and final stages of this work. 

\ \\$^*$ Electronic address: mahdi@ipm.ir

\end{document}